# SnapMode: An Intelligent and Distributed Large-Scale Fashion Image Retrieval Platform Based On BigData and Deep Generative Adversarial Network Technologies

Narges Norouzi*,[a], Reza Azmi[b], Maral Zarvani[c], Sara Saberi[d]

[a,b,c,d] Department of Computer Science
University of Alzahra
Tehran, Iran

## Abstract

Fashion is now among the largest industries worldwide, for it represents the human history and helps tell the world's story. As a result of the Fourth Industrial Revolution, the Internet has become an increasingly important source of fashion information. However, with a growing number of webpages and social data, it is nearly impossible for humans to manually catch up with the ongoing evolution and the continuously variable content in this domain. The proper management and exploitation of big data can pave the way for the substantial growth of the global economy as well as the citizen satisfaction. Therefore, computer scientists have found it challenging to handle e-commerce fashion websites by using big data and machine learning technologies. This paper first proposes a scalable focused Web Crawler engine based on the distributed computing platforms to extract and process fashion data on e-commerce websites. The role of the proposed platform is then described in developing a disentangled feature extraction method by employing deep convolutional generative adversarial networks (DCGANs) for content-based image indexing and retrieval. Finally, the state-of-the-art solutions are compared, and the results of the proposed approach are analyzed on a standard dataset. For the real-life implementation of the proposed solution, a Web-based application is developed on Apache Storm, Kafka, Solr, and Milvus platforms to create a fashion search engine called SnapMode[1].

## 1 Introduction

Now a day, fashion has occupied a large part of people's daily lives. In modern society, fashion has had a major impact on every aspect of social life so that the global fashion e-commerce market is expected to grow from $549.55 billion in 2020 to $ 672.71 billion in 2023.
The developing ubiquity of online media and the thriving of web based business has produced enormous amount of cross-media fashion data, for example, street information shared by clients, runway show information delivered by fashion brands and product data given by e-commerce sites, providing a rich and complex type of multimedia content. consequently, understanding and analyzing the semantics of massive cross-media fashion data using deep learning and image processing methods is one of the critical business analytics and technology tool for modernizing fashion industry [1].

On the other side, with an increasing amount of fashion web pages and social data, it is practically impossible for users to manually monitor of developing and ceaselessly changing content in this area. So, the proper management and exploitation of big data can pave the way for an important growth of the world economy and the citizen satisfaction [2].

---

* Corresponding author at: Department of Computer Science, University of Alzahra, Tehran, Iran
Email Addresses: n.norozi@alzahra.ac.ir (N.Norouzi), azmi@alzahra.ac.ir (R.Azmi), m.zarvani@student.alzahra.ac.ir (M.Zarvani), s.saberimoghadam@student.alzahra.ac.ir (S.Saberi)

[1] SnapMode. Sunday, 10 October 2021; Available from: https://www.snapmode.ir

One of the important communication channel to access, search and retrieve information from innumerable web sources, is a search engine that effectiveness of it extremely depends on web crawler and crawling technique used by them [3, 4]. Hundreds of web pages should be indexed in every second and facilitated users to search and navigate accordingly. So the web crawlers play a significant role in search engine systems. They have to handle a massive and ever-growing amount of data. Typically, such data cannot be processed using single node infrastructure.

In fact, big data processing is an integral part of web crawlers at dealing with billions of data. This gives rise to the need for real-time big data processing to be integrated with the search engines. In other words, With the explosive development of network scale, the single node web crawler that depends on the single machine processing capacity cannot fulfill the need of quickly acquiring large amount of resources [5]. This challenge is exacerbated when, in addition to text, there is a need for content-based image processing. Distributing the crawling process among multiple nodes can distribute processing to reduce the analysis of web page. Each crawler in the system acts as separate entity and does its own indexing [6,7].

Big data computing can be divided into two paradigms: batch-oriented computing and real-time oriented computing (or stream computing). Under the batch processing model, a set of data is collected over time, then fed into an analytics system. Apache Hadoop [8] is an example of batch-oriented computing. contrastingly, stream computing involves continual input and outcome of data. It has been used in many applications such as Tweeter Storm [9], Yahoo S4 [10] and Microsoft Time Stream [11]. In fact, big data stream processing provides real-time computing, high throughput distributed messages, and low-latency processing with massively parallel processing architectures to gain valuable knowledge from big data. The most significant techniques and technologies of this paradigm are discussed in [12,13]. Given the above explanation, it is clear that web crawling is a stream processing issue.

Scrapy, Nutch [14], HTTrack [15], YaCy [16] and stormCrawler [17], are currently popular web crawlers to crawl URLs distributedly. Although the existing distributed web crawler framework is still exploring faster speeds and more robust anti-crawling capabilities, the challenge of real-time processing and indexing of image data remains unresolved. Therefore, it is essential to design a distributed Web crawler platform that can extract disentangled features from images and index them for content-based image retrieval applications.

This paper proposes a comprehensive architecture for a distributed Web crawler platform based on open-source big data tools. The proposed architecture is modular in terms of different analytical engines, and data storage systems can easily be plugged in. Moreover, the established big data techniques can help conduct real-time data analysis, faster data integration, and low-latency processing with high throughput [18]. The proposed solution implemented and extensively assessed in the fashion domain. It can successfully be extended to other domains, such as furniture and jewelry.

In particular, this paper makes the following contributions:

- It presents a distributed stream processing platform for the real-time analysis of fashion data based on open-source big data analysis tools including Apache Storm [19], Apache Kafka [20], Apache Solr [21], and Apache Zookeeper [22].

- It proposes a disentangled feature extraction method through a deep convolutional generative adversarial networks (DCGANs) for content-based image indexing and retrieval.

- It provides a solution to power embedding similarity search through the Milvus engine [23].

The rest of this paper is organized as follows. Section 2 presents the state-of-the-art methods in the domain of stream processing engines, and the research background whereas Section 3 reviews related works. After that, Section 4 describes the proposed distributed crawler platform and the implementation details of GAN-based deep neural network models for the unsupervised extraction of visual features. The experimental results are then reported in Section 5. Finally, Section 6 draws the research conclusion and suggests future directions.

# 2   Background

In this section, we provide some background knowledge of stream processing engines, focusing on Apache Storm and Apache Kafka. We also provide a brief overview of Full-text Search Engines that are used to search large volumes of text. Finally, we introduce the Vector Search Engines that have been built to power embedding similarity search and Artificial Intelligence applications.

## 2.1 Stream Processing Engines (SPEs)

Data Stream is data, that persistently moves from a source to a destination to be processed and analyzed in near real-time. The data stream can be produced in at various rates and message size, depending on the application. The industries, for example Internet of Thing and social media, that heavily depend on real time processing have made streaming data even more important [24].
Stream processing often involves several tasks on the stream data which can be performed in parallel or/and serially manner. According to [24] stream processing engines can be categorized into three generations.
The first generation of SPEs were actually event-based systems and mainly consisted of databases and rule engines. Using these systems, users could define actions that would be performed when certain conditions were met. PostgreSQL database is a well-known example of this generation. But the second generation of these systems were specifically implemented for the purpose of processing data stream. Hence, they had a significant improvement in stream processing compared to the first generation. A Directed Acyclic Graph (DAG) was used by Aurora to represent streaming process but was limited to a single node whereas Borealis [25] implemented it is a distributed cluster-based system. Third generation of SPEs were the first systems designed to process a distributed stream source. The frameworks from this generation include Apache Storm, Apache Flink, Apache Beam [26], Apache S4, MillWheel [27], Amazon Kinesis [28], and IBM InfoSphere [29].
Apache Storm, has emerged as one of the earlier systems for distributed real-time stream processing and has been widely adopted. The main advantages of this framework can be mentioned as follows:
- Very low latency, true streaming, mature and high throughput: The Storm software comes with the latency of just a few milliseconds on micro-batch processing, which obviously makes it a reliable data processor. Reliability is a factor that helps Storm stand out as a real-time data processing system.
- Excellent for complicated streaming use cases: The standard configuration of Storm makes it fit instantly for production among small and medium enterprises along with big-sized organizations.
In Apache Storm framework, each store cluster consists of Nimbus node(s) and Supervisor nodes(s). Distributing code around the cluster, assigning tasks to machines, and monitoring for failures are Nimbus node(s) responsibilities. Each supervisor can run workers which are separate JVM processes on its node. Each worker process executes a subset of a topology which is a network of spouts and bolts. Spout is the entry point and the source of streams in a storm topology. A spout connects to the data source, gets data stream, converts the actual data into stream of tuples, emits them to bolts for actual processing. Bolt contains the actual processing logic and can emit streams too for more processing downstream by other bolts or can save data for persistent storage. Important point to note is, Apache Storm uses Apache ZooKeeper [22] to manage the cluster state. All coordination between Nimbus and the Supervisors such as message acknowledgements, processing status, etc. is done through a Zookeeper cluster.

## 2.2 Message Oriented Middlewares (MOMs)

Since stream processing engines often keep internal states in main memory, which is fast but unreliable, the generated stream data will become unavailable after being processed. Given these conditions, if the stream processing engine consumes a message from the stream source but fails before completely processing it, that message is all but lost. Therefore, fault tolerance is a critical requirement in distributed stream processing engines because failures can happen at any time and at any worker in a distributed environment.  In order to address this issue, the concept of Message Oriented Middleware was introduced. MOMs are software infrastructures which support the sending and receiving of messages between component information systems in an enterprise's distributed system. In fact, they are an intermediate layer between the stream source, and the SPE that the stream source publishes messages to the MOMs and an SPE subscribes and consumes message from it. In an environment equipped with MOM-based communication, messages are usually sent and received asynchronously. Using this communications, senders and receivers are never aware of each other. in other words, applications in this environment are abstractly decoupled. Instead, they communicate with each other via sending and receiving messages to and from the messaging system [30].

Apache Kafka [20] is a distributed data streaming platform which is designed for message-oriented middleware and used most often for streaming data in real-time into other systems. In fact, it is a middle layer to decouple your real-time data pipelines. Hence, Kafka can be used to feed fast lane operational data systems like Storm.

Kafka combines two messaging models, queuing and publish-subscribe, to provide the key benefits of each to consumers and allows multiple independent applications reading from data streams to work independently at their own rate.

**2.3 Full-text Search Engines**

In a full-*text* search, a *search engine* examines all of the words in every stored document as it tries to match search criteria giving by the user.

Apache Solr [21] is one of the popular and faster open source enterprise search platform built on top of Apache Lucene. It is packed with features like load-balancing queries, automated functions, centralized configuration, distributed instant indexing and scale-ready infrastructure. Solr takes in unstructured/structured data from different sources, indexes and stores it, and finally makes it available for search in near real-time. Solr is also provides analytical capabilities as well, enabling you to do faceted product search, log/security event aggregation and social media analysis. Also, it can work with large amounts of data in what has traditionally been called master-slave mode, but it allows further scaling via clusters in SolrCloud mode.

**2.4 Vector Search Engines (VSEs)**

Vector-Searching enables a large spectrum of use cases which are impossible with traditional full-text search engines. A pure full-text search tries to match occurrences of terms in a set of documents. On the other hand, a vector search engine ranks each object by how close it is to another.

In machine learning, objects and concepts are often represented as a set of continuous numbers, also known as vector embeddings. This very straight method makes it possible to translate the similarity between objects embedded as a vector space. This means when we represent images or pieces of text as vector embeddings, their semantic similarity is measured by how close their vectors are in the vector space. Hence, what we want to look at is the distance between vectors of the objects.

Milvus [23] is a cloud-native, open-source vector database built to manage embedding vectors generated by machine learning models and neural networks. It extends the capabilities of best-in-class approximate nearest neighbor (ANN) search libraries and features on-demand scalability, and high availability. The goal of Milvus is to simplify unstructured data management and provide a consistent experience across different deployment environments [23]. Milvus is widely used in different scenarios, such as image, video and audio similarity search, DNA sequence classification, question answering and recommender systems.

**2.5 Content-based Image retrievals (CBIRs)**

Content Based Image Retrieval (CBIR) is a method which uses features of image to search from large image dataset according to the user's request in the form of query image. For any CBIR system, there are two main step: finding the most important features of each image so that images can be described as feature vectors and calculating distances between images for similarity. Therefore, Effective feature representation and similarity measures are very crucial to the retrieval performance of CBIR [31]. In early days, various hand designed feature descriptors have been investigated based on the visual cues such as color, texture, shape, etc. that represent the images. However, the deep learning has emerged as a dominating alternative of hand-designed feature engineering from a decade. In Fact, one of the most interesting applications of deep learning in recent years is creating feature representations of images as vectors. Convolutional Neural Networks (CNNs) have widely been used for this purpose and show great promise. Another main approach for creating vector representation is Generative Adversarial Networks (GANs) [32].

## 2.6 Generative Adversarial Networks (GANs)

Generative Adversarial Networks(GANs), are an approach to generative modeling using deep learning methods, such as convolutional neural networks. Generative modeling is an unsupervised learning task in machine learning that involves automatically discovering and learning the patterns in input data in such a way that the model can be used to generate or output new examples that plausibly could have been drawn from the original dataset. GANs consist of two main parts: generators and discriminators. The generator model produces synthetic examples from random noise sampled using a distribution, which along with real examples from a training data set are fed to the discriminator, which attempts to distinguish between the two. Both the generator and discriminator improve in their respective abilities until the discriminator is unable to tell the real examples from the synthesized examples with better than the 50% accuracy expected of chance. GANs have successfully been used in various applications such as generating visually realistic images, style transfer, image-to-image translation (CycleGAN), creating high-resolution images from low resolution samples (SRGAN), image generation from text (StackGAN), learning to discover relationships between different domains such as fashion items (DiscoGAN), transferring facial makeup from a reference image (BeautyGan), and other applications reviewed in survey [33].

## 3 Related Work

Generally speaking, a search engine operates through three main processes: scanning (crawling) the Web, indexing, and ranking (sorting) the results obtained from a crawler, and visualizing results [3]. Distributed web crawler is a distributed computing technique which crawls Web resources on the Internet according to some rules and provides the obtained network information to search engine. Therefore, it is an essential part of search engines. Since the output of search engines is a list of snapshot of web pages, it is impossible to find any specific structured data using search engines. We have to use a special web crawler to collect specific and structured data called Focused Web Crawler.

Focused web crawler, sometimes called vertical or topical web crawler, is a tool that selectively crawl pages related to predefined topics and plays an important role in Information Management Systems. A large variety of focused crawlers has been created, aiming to extract specific information from the web. Several studies show the application of the focused approach in different contexts: extracting hidden information from the e-commerce websites [34], Deep Web [35] or for discovering, annotating, and classifying biomedical information [36]. In the recent years, some researchers optimized the precision of focused web crawling results by implementing different approach.

A recent study has suggested a new approach for web crawling, based on reinforcement learning, which considers the distance between two pages as the average number of clicks to reach one page from the other [37].

In [38], the authors propose an approach to extracts Term Frequency (TF) based features and optimizes keyword weight using the Stochastic Gradient Descent (SGD) algorithm.

The proposed crawler at [39] prioritizes the URLs in the queue, which allows the search engine to visit important pages first. So by adopting the Focused Crawling strategy, the proposed crawler limits the Web exploration to the pages that are relevant to some predefined topics.

A classic approach to semantic search leverages on domain ontologies [40], in order to model the semantics of terms and entities belonging to a given domain, by indicating their attributes and relationships.

In the focused web crawlers, one of the most important problems to overcome is to increase the speed of crawling politely. A multi-threaded crawler is proposed [41], which can speed up crawling, however it is detected as an impolite crawler if used to collect data from a web site. Other researchers developed distributed crawler [42,43], But each of them has different limitations for use in real issues.

Scrapy, Nutch [14], HTTrack [15], JSpider, YaCy [16] and stormCrawler [17], are currently popular web crawlers to crawl URLs distributedly. Although the existing distributed web crawler framework is still exploring faster speeds and more robust anti-crawling capabilities, the challenge of real-time processing and indexing of image data remains unresolved.

To overcome the above challenges, this paper proposes a distributed focused web crawler which uses big data tools and deep GANs to extract disentangled features from images and index them for content-based image retrieval applications.

# 4  Proposed Framework

This section proposes an intelligent distributed focused Web crawler framework called SnapMode [49] to extract and process fashion data from the e-commerce websites by integrating a collection of suitable technologies from the existing state-of-the-art methods. The proposed framework is described in terms of how it handles the challenges of stream data analysis on fashion websites discussed previously.

## 4.2  High-Level System Architecture

The proposed framework addresses the complex task of integrating data from different sources and large-scale stream processing through open-source Apache Solr and Apache Kafka. Figure 1 demonstrates the high-level architecture of the proposed framework that consists of six main layers called (i) data storage layer, (ii) data extraction and stream process layer, (iii) messaging layer (iv) stream analytics layer, (v) business layer, and (vi) presentation layer. In the data extraction and stream process layer, the aggregated data are processed through batch or stream distributed processing engines such as Storm and Spark. The messaging layer is responsible for handling the streamed data generated by the first layer and containing image URLs. Furthermore, the stream analytics layer normally incorporates the image processing engines responsible for the real-time analysis of image data, whereas the data storage layer is responsible for storing the real-time data analysis output. The data extracted from the stream process layer, which is in the textual format, are stored in the Solr database. The image embedding vectors generated by the analytics layer are then stored in the Milvus database. The business layer contains certain objects that execute the business functions and handle the exchange of information between a database and a user interface. This layer used FastApi, which is a modern high-performance Web framework, to develop APIs through Python. Finally, the presentation layer is an interaction layer required for end users, internal employees, and business customers to access the application. The Vue.js framework was employed to implement Web applications. It is an open-source progressive JavaScript framework for developing user interfaces and single-page applications.

Figure 2 presents an overview of the system architecture consisting of two main subsystems and describes how they communicate with each other through the Kafka framework used as a message broker and an intermediary. The crawler subsystem consists of a storm topology that has two main functions, (i) extracting product information from the crawled webpages and (ii) sending image URLs to the image processing subsystem. Since image processing is a very time-consuming process that requires a sophisticated topology, feature extraction modules are considered separately from the text processing ones. The separation of these two subsystems would result in the process of extracting and updating different pieces of information required by end users, e.g., price, discount rate, number of non-delayed stocks.

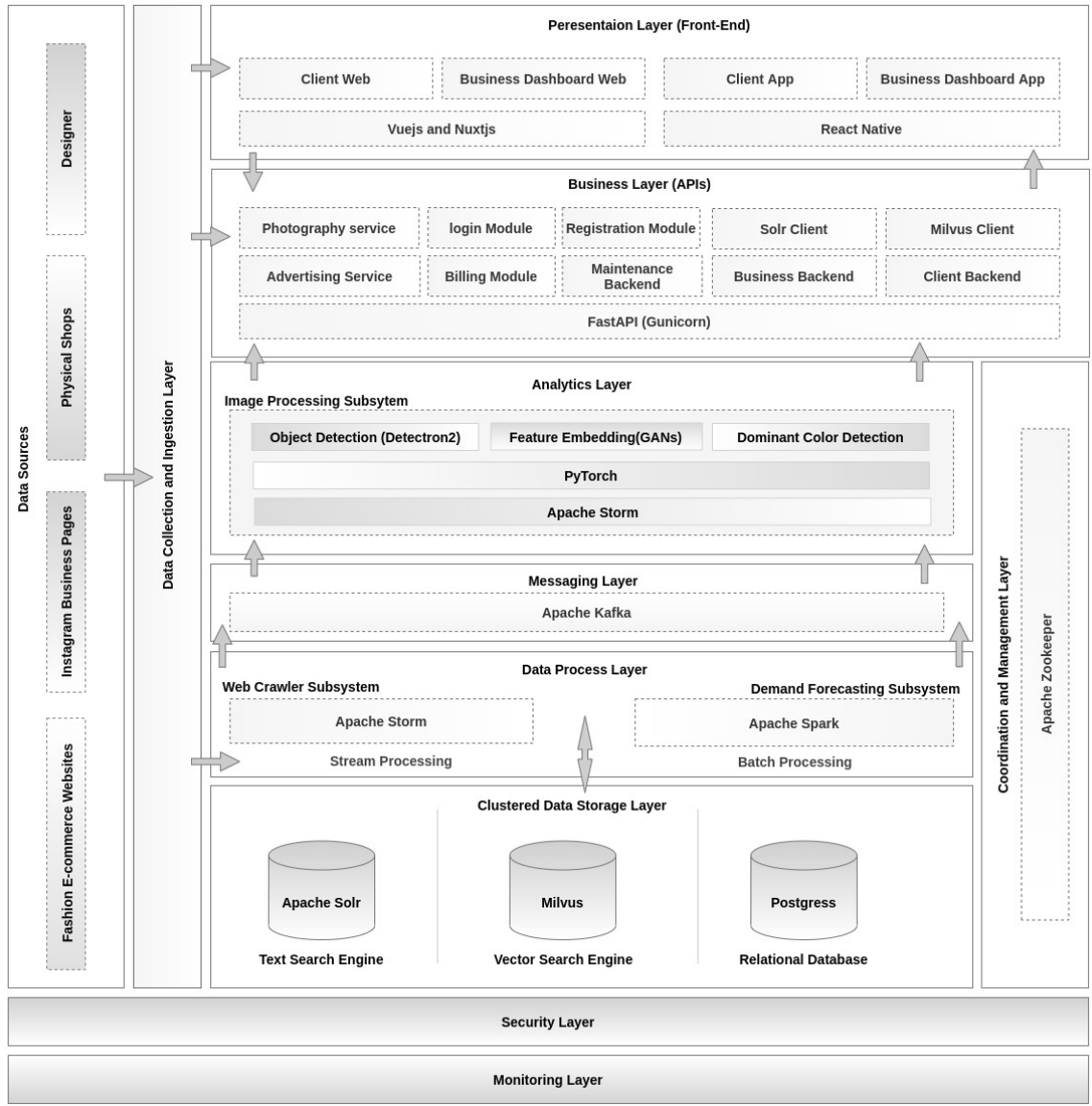

Figure 1: Overall view of system architecture

The Web crawler topology extracts product information including name, price, brand, description, links, and image URLs and stores the outputs in the Solar database. The last bolt of this topology acts as a Kafka producer and publishes image URLs to a topic. At the same time, the spout of image processing topology acts as a Kafka consumer and subscribes the image URLs from the Kafka topic to extract feature vectors. The output of this subsystem is stored in the Milvus database. Each of these topologies is described in detail below.

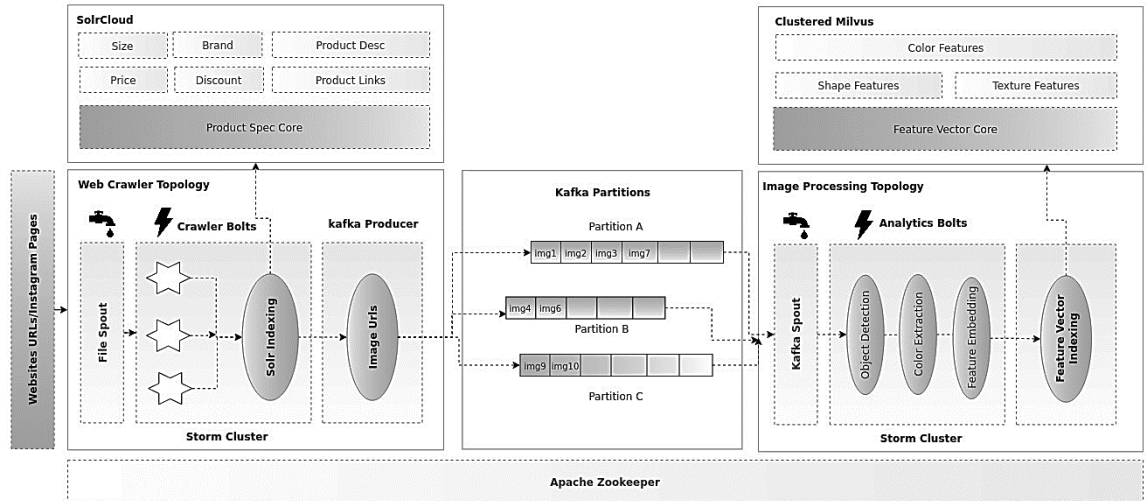

Figure 2: Overview of two main subsystems: (i) Web Crawler and (ii) Image Processing

### 4.3 Crawler Subsystem

Figure 3 depicts the internal components of the crawler subsystem and the way in which they interact with each other. This subsystem is modular and consists of two core modules called (i) the crawler topology and (ii) the topology scheduler. They provide the basic building blocks of a Web crawler such as fetching, parsing, and indexing.

When e-commerce businesses register their crawl requests in the system, the topology scheduler receives the website addresses, extracts the URLs of product webpages, and stores them in the relevant Solr core. After that, it builds a topology with specific configurations and queues it for execution. This module is also responsible for executing, activating, and deactivating topologies. The crawler topology consists of two spouts and 10 bolts. The spouts read the URL tuples from a Solr core and emit them to the URL partitioner bolt, in which the target URLs are extracted and emitted to the fetcher bolt to fetch the contents of the product pages. After initial clearing, the fetcher bolt emits the page contents in parallel to the information extraction bolts. Each of these bolts, as the name implies, is responsible for extracting specific information from product webpages. For instance, the price bolt extracts the price information. The outputs generated from the above bolts are then emitted to the indexing bolt, which is responsible for storing information in the Solr database. Furthermore, the Kafka producer bolt publishes the image URLs in the Kafka topic.

### 4.4 Image Analytics Subsystem

According to Figure 4, an image retrieval system basically consists of a feature extraction model, vector database, and a similarity matching module. When a user inserts an inputs in the form of a query image or clicks on an image, its feature vectors are extracted and compared with the features of images stored in the database. After similarity matching, images relevant to a query image are retrieved. It is now challenging to search for an image in a large number of image datasets based on its semantics with high accuracy and performance.

The image processing subsystem is actually a content-based image retrieval system designed for large-scale stream image processing and indexing on a distributed environment. Figure 5 demonstrates an overview of this subsystem and its constituent modules.

In this subsystem, a Storm topology is defined to extract image features through DCGANs, in which a Kafka spout is used as a consumer that subscribes the image URLs from Kafka partitions and emits them to the first bolt that is responsible for object detection. Clothing object detection includes detecting the specific regions of the clothing objects existing in a given image. For example, given an image of an individual wearing a full outfit, clothing object detection involves the prediction of bounding boxes that would capture the distinct clothing articles such as a shirt, pants, and shoes. Instead of using an entire image as a query, this spout employs an object detection algorithm as a preprocessing step. Therefore, it will be possible to retrieve the results with higher precision, for the search process looks for only within a limited space of pictures from the same class. The DeepFashion2 dataset [44] was used for the clothing object detection task. It contains 491K diverse images of 13 popular clothing categories from both commercial shopping stores and consumers. For object detection, the YOLOv4 [45], which is based on the Darknet-53 model, was trained for the MS COCO [46] dataset. In fact, a dense layer of 80 labels was removed, and a new dense layer of 13 labels representing clothing categories was added. The model was then trained through the DeepFashion2 dataset, and the pre-trained features were transferred into a fashion domain by using the transfer learning power. When the regions of interest are extracted, the object detection bolt emits them to a feature extraction bolt that uses a DCGAN for image representation. Finally, the extracted output vectors are stored in Milvus after normalization.

As discussed earlier, the feature extraction module is the most critical component of image retrieval systems, for the accuracy of results depends on the quality of the extracted features. Over the past years, several machine learning algorithms have been developed to extract features from data. Deep learning models such as CNNs yield successful results for image representation; however, as the evaluation process will indicate, the results are not accurate enough to extract effective features within the fashion domain. In recent years, Generative Adversarial Networks have yielded significant outputs in realistic and high-fashion clothing generation. Hence, GANs were employed in this study to fine-grain feature representation.

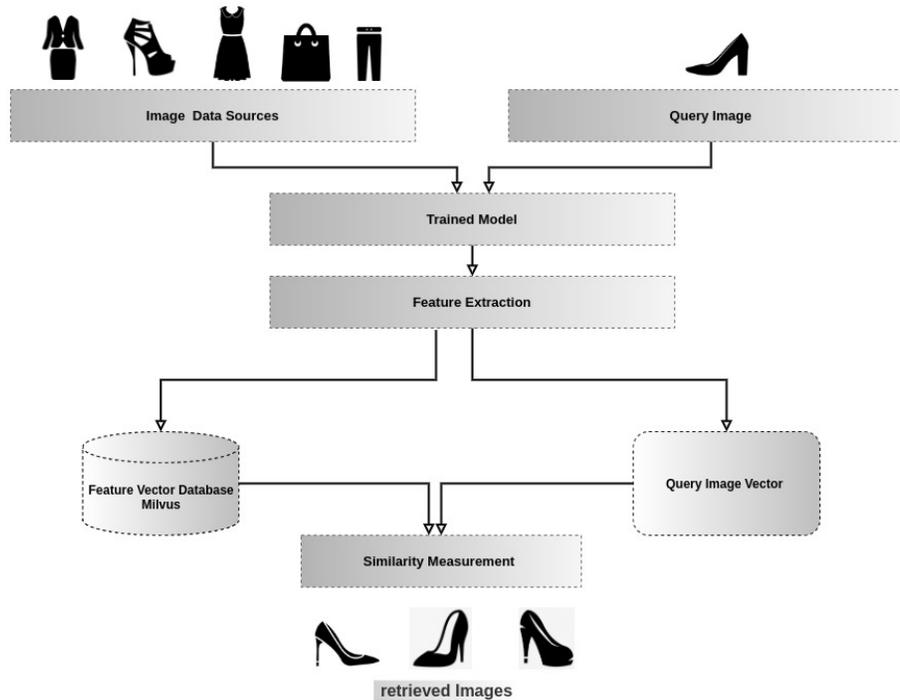

Figure 4: Overview of Content-based Image Retrieval System

### 4.5 Vanilla GAN

GANs are a powerful class of neural networks that are used for unsupervised learning. It was developed and introduced by Ian J. Goodfellow in 2014 [32]. GANs can be broken down into three modules: (i) Generative: To learn a generative model, which describes how data is generated in terms of a probabilistic model. (ii) Adversarial: The training of a model is done in an adversarial setting. (iii) Networks: Use deep neural networks as the artificial intelligence algorithms for training purpose.

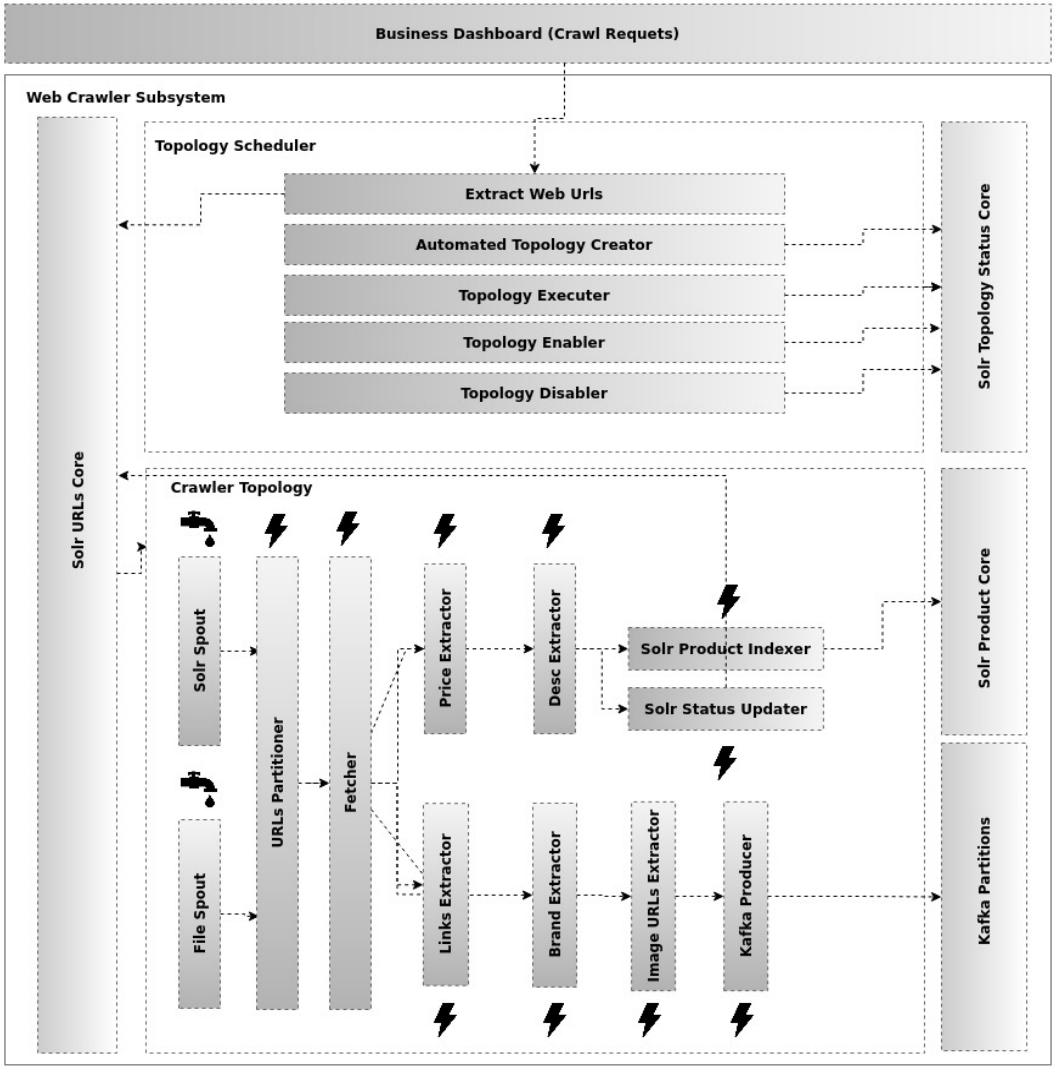

Figure 3: Overview of Crawler Subsystem

The generator generates fake samples of data (for example, image) and tries to fool the discriminator. The discriminator, on the other hand, tries to distinguish between the real and fake samples. The generator and the discriminator are both Neural Networks and they both run in competition with each other in the training phase. The steps are repeated several times and in this, the generator and discriminator get better and better in their respective jobs after each repetition. The working can be visualized by the diagram given Figure 6.

As described in [47] discriminator D and generator G play a minimax game in which D tries to maximize the probability it correctly classifies reals and fakes (log D(x)), an G tries to minimize the probability that D will predict its outputs are fake (log(1−D(G(z)))). From the paper [32], the GAN loss function is:

$$\min_G \max_D V(D, G) = E_{x \sim p_{data}(x)}[\log D(x)] + E_{z \sim p_z(z)}[\log(1 - D(G(z)))] \quad (1)$$

where, *G* is Generator and *D* is Discriminator, $P_{data}(x)$ and $P(z)$ indicate real data and noise distribution, respectively. $x = \{x_1, x_2, ..., x_n\}$ is a sample form $P_{data}(x)$ and $z \in \mathbb{R}_z$ is a sample from *P(z)*, *D(x)* and *G(z)* are Discriminator and Generator networks.

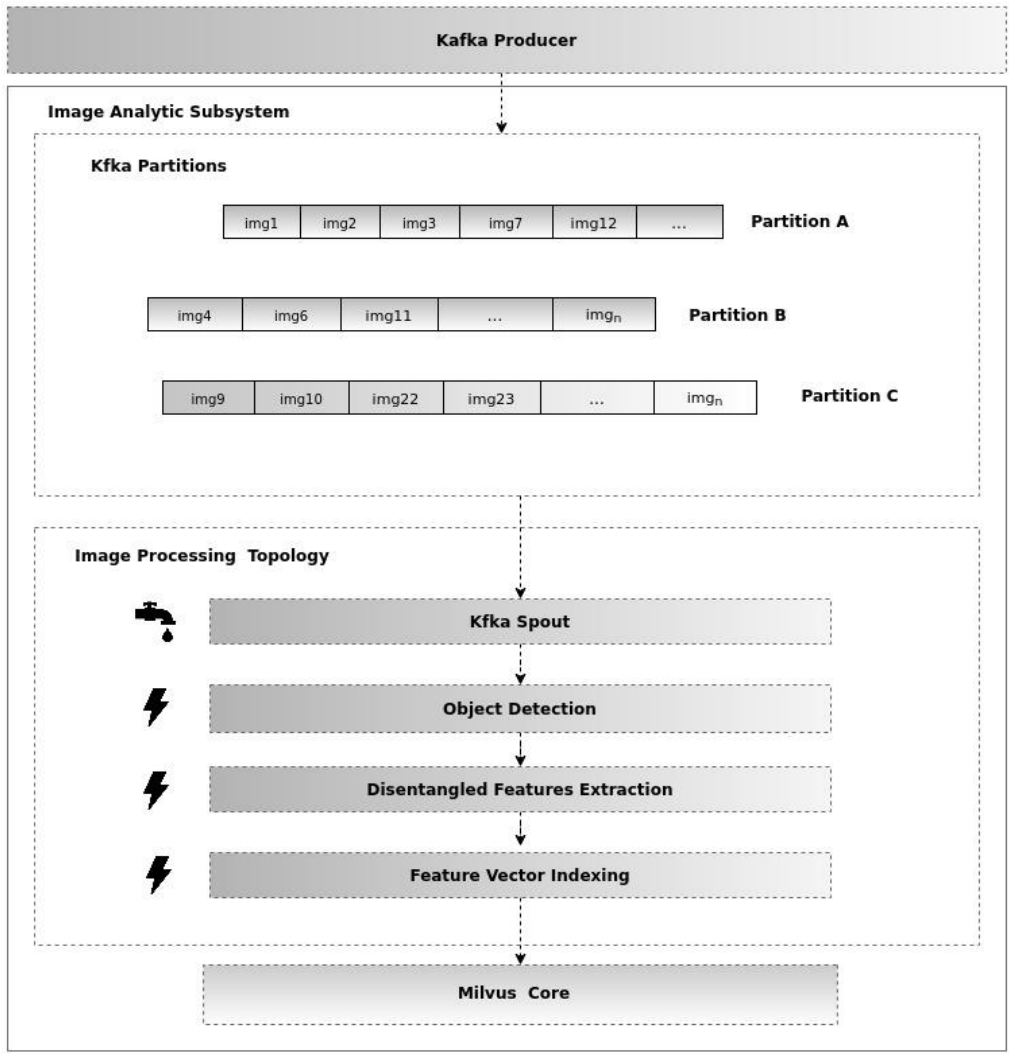

Figure 5: Overview of Image Processing Subsystem

### 4.6 Deep Convolutional Generative Adversarial Networks (DCGANs)

By definition, a DCGAN is a direct extension of the GAN described above; however, it explicitly uses convolutional and convolutional-transpose layers in the discriminator and the generator, respectively [47]. The discriminator consists of the strided convolutional layers, batch norm layers, and LeakyReLU activations. The input is a 3×64×64 image, whereas the output is a scalar probability that the input comes from the real data distribution. The generator consists of convolutional-transpose layers, batch norm layers, and ReLU activations. Its input is a latent vector (z) drawn from a standard normal distribution, whereas its output is a 3×64×64 RGB image. The strided convolutional-transpose layers allow the latent vector to be transformed into a volume with the same shape as an image. Figure **7** demonstrates the generator and discriminator networks of the DCGAN architecture.

The transferring knowledge of the pre-trained and fine-tuned networks is widely used in the techniques for enhancing the performance of discriminative models. Hence, the pre-trained DCGAN model was fined-tuned with images of the DeepFashion2 dataset to develop a new image generator and a discriminator. In the process of fine-tuning the model, the learning rate is set to 0.0002, whereas the number of mini-batch repetitions is set to 128 on those reported by [47]. The selected values for the learning rate and mini-batch repetition increase the stability and speed during the training process.

As mentioned above, in this model, the discriminator D takes an image with tree-color channels and 64×64 pixels in size and yields a binary prediction as fake or real. Instead of the binary prediction, the discriminator was employed to extract a 1×8192 feature vector after the trained model was saved.

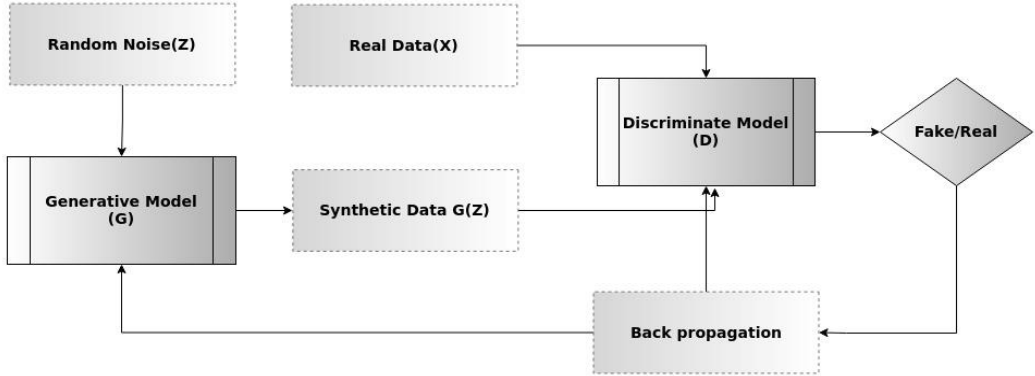

Figure 6: Vanilla GAN

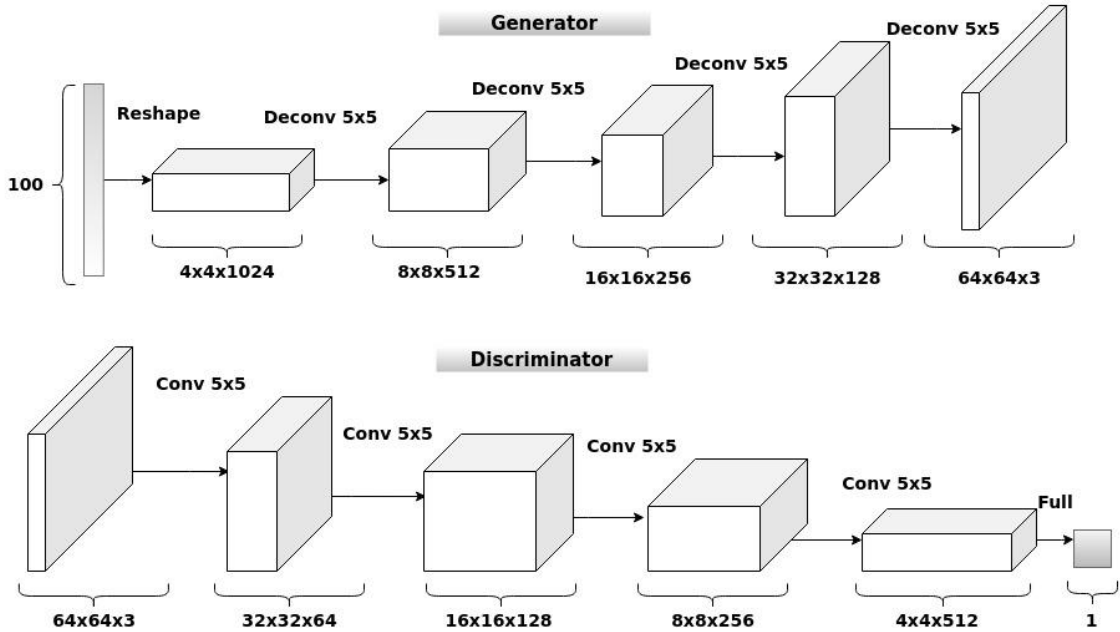

Figure 7: The generator and discriminator networks in the DCGAN model

## 5 Experimental Results

A number of experiments were conducted to evaluate the proposed framework in the image processing subsystem. This section presents the corresponding experiment materials. Subsection 5.1 illustrates the experimental environment and describes the datasets used in this study, whereas Subsection 5.2 indicates how to fine-tune the state-of-the-art deep learning networks. In addition, Subsection 5.3 reports the comparative results of the proposed method. Finally, Subsection 5.4 introduces SnapMode.

## 5.1 Datasets and Experimental Setup

For the performance analysis of the proposed method, the experiments were conducted on a Linux-based OS (Ubuntu 18.04) running on an HPE ProLiant DL380 Gen10 server consisting of a 28-core Intel Xeon Silver 4100 CPU with 128 GB of RAM and an NVIDIA GeForce GTX 1080-Ti GPU. To ensure that the proposed framework operated efficiently with different images, two datasets were employed for testing. DeepFashion2 [44] is a comprehensive fashion dataset containing 491K diverse images of 13 popular clothing categories from both commercial shopping stores and consumers. DeepFashion2 In-Shop Retrieval contains 7,982 clothing items with 52,712 images. 3,997 items are for training (25,882 images) and 3,985 items are for testing (28,760 images). The test set is composed of a query set and a gallery set, where query set contains 14,218 images of 3,985 items and gallery set contains 12,612 images of 3,985 items.

In addition to evaluating the proposed method through DeepFashion2, we used our own dataset collected from Iranian e-commerce websites (i.e., https://www.digikala.com/, https://www.digistyle.com/, https://www.banimode.com/, https://www.modiseh.com/, https://banistyle.com/, https://tagmond.com/, https://fiza.ir/, https://hamechiinjast.com/author/modirhame/, https://epasazh.com/, https://aassttiinn.com/). The final dataset consisted of 639,716 images crawled from 730 websites. In the end, training and inference were performed through the open-source TensorFlow platform.

## 5.2 State-of-the-Art CNNs

This study employed and evaluated several popular CNNs models such as ResNet101V2, ResNet152V2, InceptionResNetV2, NASNetLarge, DensNet201, MobileNetV2, and VGG19. The baseline models trained for the ImageNet dataset classification were also fine-tuned on Deepfashion2. In fact, the models were fine-tuned by replacing the Softmax classification layer of a pre-trained baseline model with a new classification layer trained to classify DeepFashion2 images initialized through the same weights of intermediate and lower layers. Therefore, the pre-trained features were transferred into fashion domains by through the transfer learning power.

## 5.3 Comparative Results

It is difficult to evaluate the unsupervised learning approaches, for it is very complicated to identify KPIs which can be employed to validate the results. Thus, there is no universal consensus that performance metrics are available for image retrieval systems. However, precision is the most common evaluation indicator for the evaluation of the machine learning methods. It can be used through the following standard definition:

precision = # of relevant items retrieved and sorted by distance / # of all retrieved items

An image search result is assumed to be relevant to a query image if the two images share the same class label. This approach is commonly used for the offline evaluation of image retrieval systems [48]. To evaluate the proposed model, 20,000 images were randomly selected from the DeepFashion2 dataset (the test category of commercial images) as query images, whereas the rest of the images were indexed by the image analytics subsystem in the Milvus database. Table 1 reports the experimental results for the proposed generative network and other popular convolutional neural networks. The test results include model size, feature extraction time, and precision rate.

According to Table 1, the increase in the precision rate is mainly due to an increase in the network depth or size (e.g., NASNetLarge, VGG19, ResNet152V2). However, since the neural network depth increases, the additional cost of inference time increases and might lead to negative user experience. Therefore, the balance between precision and inference time is very important in search engines. The proposed model outperformed the other popular models in term of precision and inference time, for it reached up to 62% with an inference time of 0.02 seconds. At the same time, the proposed discriminator model, based on transfer learning and trained with the DCGAN, yielded higher precision rates than the other heavy pre-trained models by tacking an average of 0.02 seconds per each query image.

Figure 8 reports the qualitative results of the proposed embedding model compared with those of the other heavily trained models.

| Models | Evaluation Metrics | | |
|---|---|---|---|
| | Size (MB) | Inference Time (Second) | Precision (%) |
| ResNet101V2 | 171 | 0.18 | 0.548 |
| ResNet152V2 | 232 | 0.25 | 0.557 |
| InceptionResNetV2 | 215 | 0.06 | 0.573 |
| NASNetLarge | 343 | 1.09 | 0.595 |
| DensNet201 | 80 | 0.15 | 0.524 |
| MobileNetV2 | 14 | 0.04 | 0.443 |
| VGG19 | 549 | 0.24 | 0.472 |
| **Proposed Model** | **16.75** | **0.02** | **0.624** |

Table 1. Performance comparison of the proposed model and the state-of-the-art CNN models

Figure 9 demonstrates some examples of retrieval results for the query images selected randomly from the research dataset. They were not seen by the model during the training process.

### 5.4 Web Application

For the real-life use of the proposed method, a Web-based application was implemented through Apache Storm, Kafka, Solr, and Milvus frameworks to develop a fashion search engine called SnapMode. The application allows users to search for the desired products on all fashion websites and see similar products, implemented through FastAPI and Vue.js, the application is now publicly available [49]. Figure 10 shows a screenshot of the working Web application with the proposed framework.

### 5.5 Conclusion

This paper proposed an end-to-end solution for a large-scale Web crawler and a search engine in the e-commerce fashion. It summarizes an overview of a distributed stream processing framework for the crawling and real-time analysis of data based on open-source big data analysis platforms such as Apache Storm, Kafka, Solr, and Zookeeper. It proposes a disentangled feature extraction method through deep convolutional generative adversarial networks (DCGANs) for content-based image indexing and retrieval and also provides a solution to power embedding similarity search in the Milvus database. Several popular deep neural network architectures were compared with the proposed network in clothing image retrieval. According to the results, the proposed model outperformed the other pre-trained models in term of precision and inference time.
We also intend to explore other GAN architectures for generating more accurate image representation and develop the system for use in other areas such as interior design and jewelry.

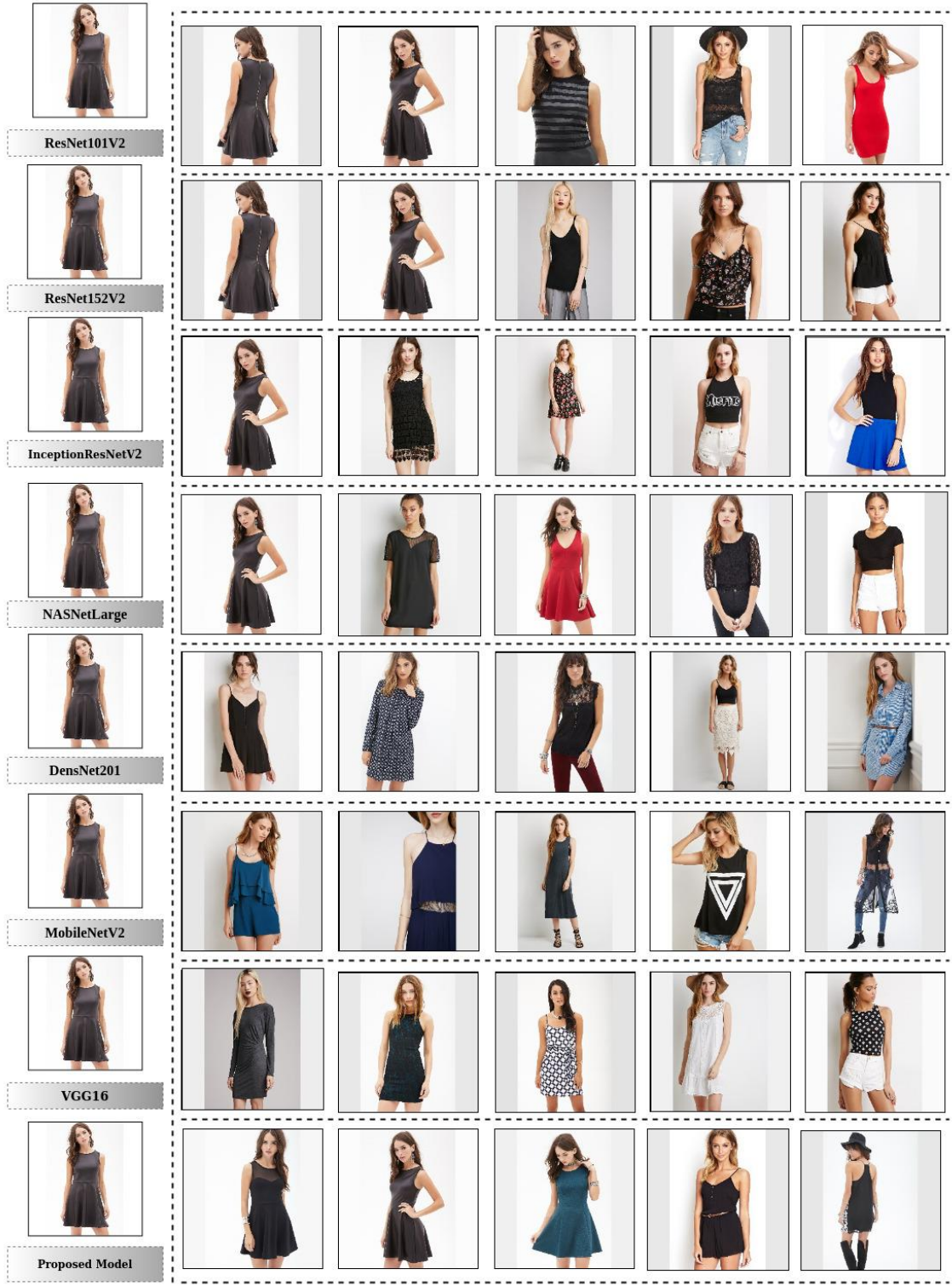

Figure 8: Qualitative results comparing our proposed model vs other state-of-the-art CNN models on DeepFashion2.

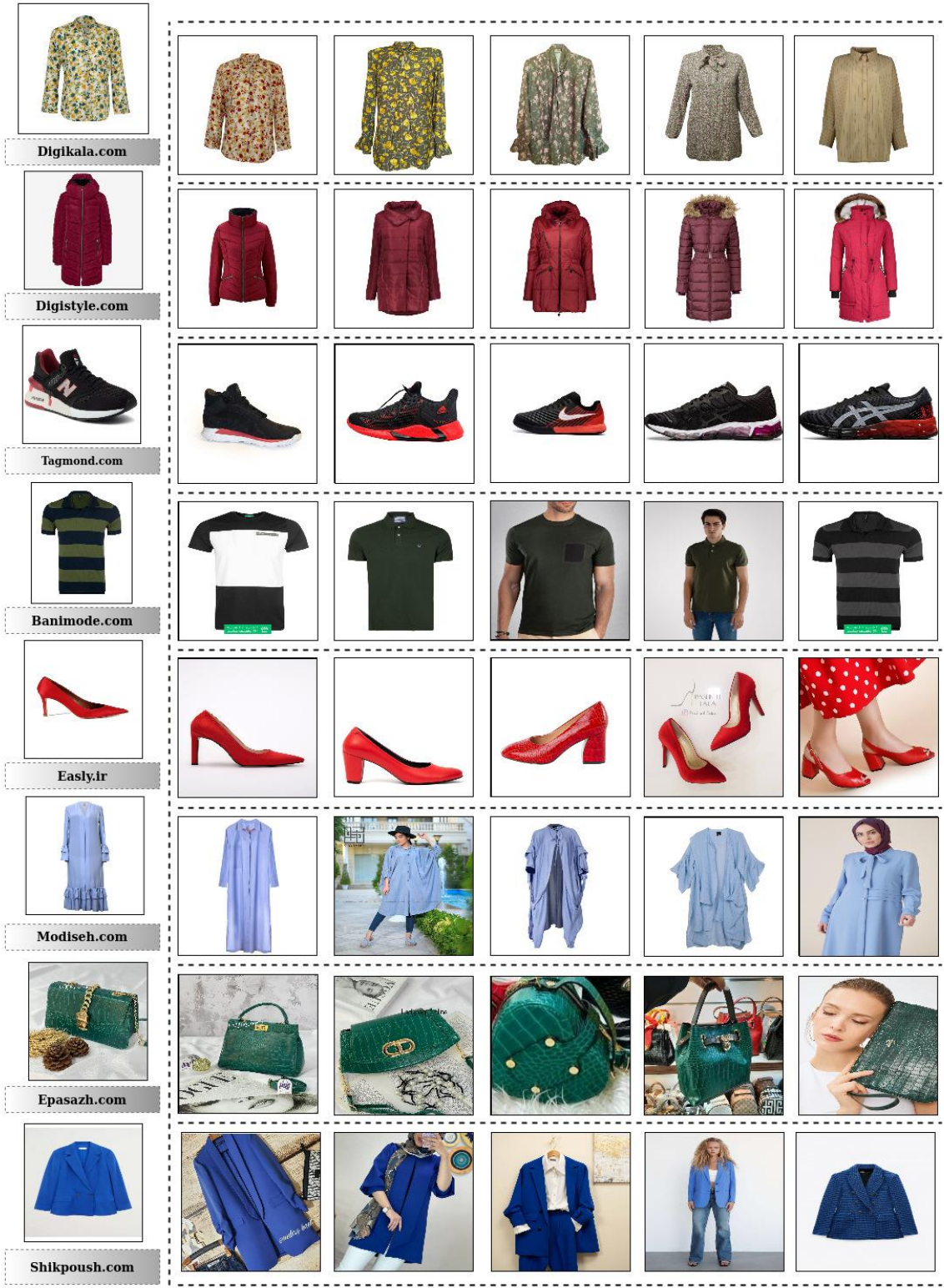

Figure 9. Sample results of the proposed model on unseen images crawled from Iranian e-commerce websites.

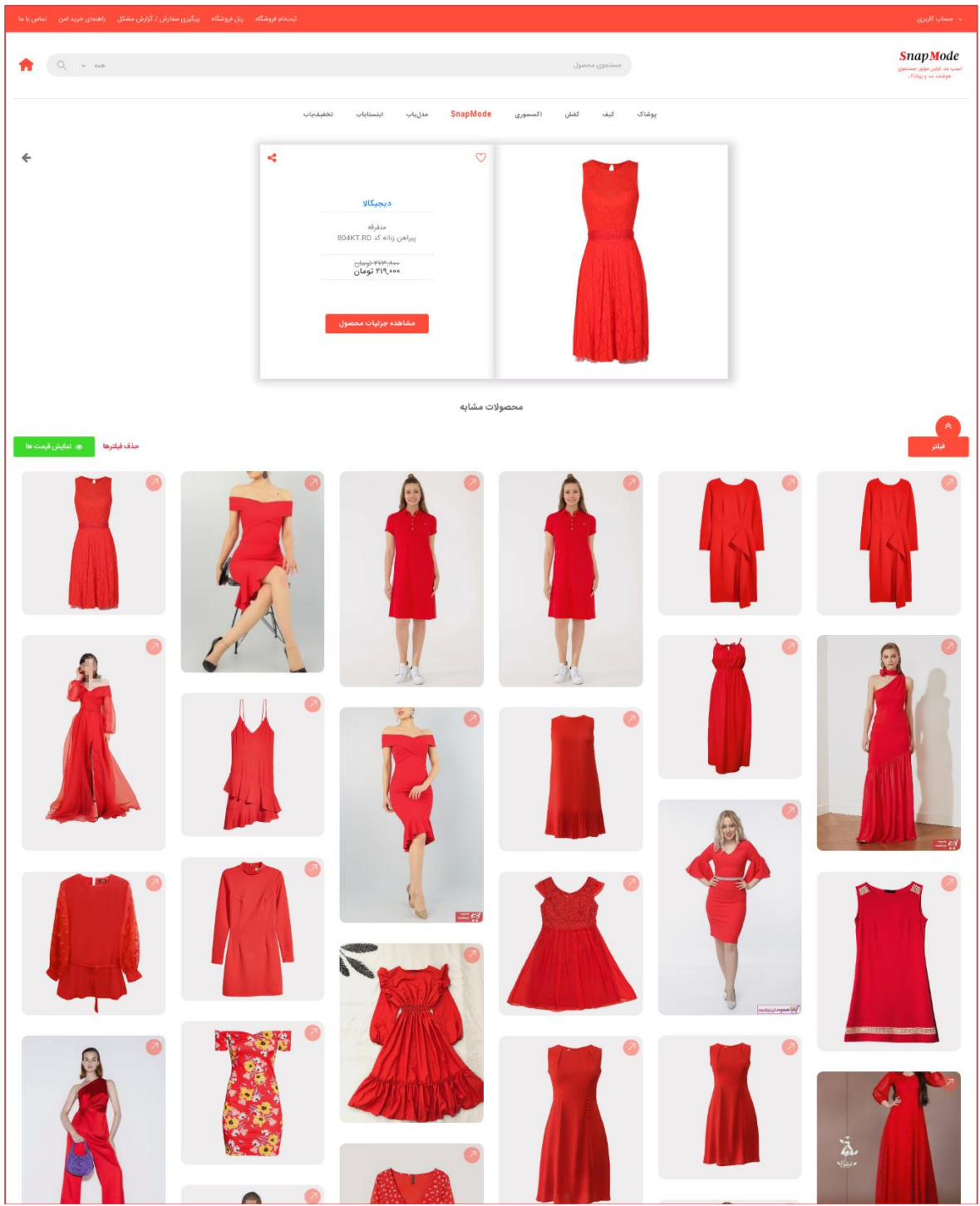

Figure 10: The sample screenshot of the proposed search engine (SnapMode) for fashion e-commerce in the Web application showing the retrieval of visually similar products.